\documentclass{article}

\usepackage{amsmath}
\usepackage{amsfonts}
\usepackage{amsmath}
\usepackage{latexsym}
\usepackage{graphicx}

\def\bea{\begin{eqnarray}}
\def\eea{\end{eqnarray}}
\def\nn{\nonumber}
\title{Radiation of the atomic chains exited by a channeling particle}
\author{V. Epp and M.A. Sosedova\\
Tomsk State Pedagogical University}
\date{}
\begin{document}
\maketitle

\begin{abstract}
Basic properties of radiation of the atomic chains excited by a channeling particle are considered. Using a very simple two-dimensional  model of a crystal lattice we have shown that the main part of this radiation is generated on the frequency of oscillations of a channeling particle between the crystal planes, shifted by the Doppler effect. Angular distribution of the  radiation of the chain of oscillating atoms is sharply peaked in the direction of the velocity of channeling particle because of coherence of the fields, produced by  individual atoms.
\end{abstract}

\section{Introduction}
Physics of channeling of the accelerated particles in crystal, is a quickly developing field of science \cite{Baryshevsky}. The effect of channeling has served as the base of development of new experimental methods of research on the crystal structure. By use of the channeling effect it is possible to study thermal oscillations and displacement of atoms in a lattice, distribution of electronic density in interatomic space of crystals and  to measure exact orientation of the crystal planes. Channeling of light particles is used for generation of intensive monochromatic X-ray radiation. Using experiments and the theories which describe the orientation effects in crystals, a new sources of X-ray and the gamma radiations consisting of accelerators and precisely oriented crystals are created.

In the course of interaction between the channeling particle and a crystal the electromagnetic radiation is generated. In the framework of classical electrodynamics one can consider this radiation as radiation from different sources -- radiation of the channeling particle oscillating between the atomic planes or around an crystal axis, radiation of the electronic gas excited by a channeling particle (weik fields of electrons) and the radiation which is produced by the excited atoms of a crystal lattice. Radiation of a channeling particle is investigated in details (see for example \cite{Lindhard, Shulga}), the experimental research of radiation of weik fields can be found in \cite{Kil}. Radiation of the atoms excited by a channeling particle is has not been studied so far. In this article we consider part of this radiation caused by oscillations of the atoms embedded at specific sites in the lattice of a crystal, leaving aside the radiation caused by quantum transitions of electrons from excited to principal states. We suppose that the conditions of implementation of classical electrodynamics are fulfilled. Namely, the  energy of excitation of atom is much more than distance between the energy levels; we consider the part of a radiation spectrum of single atom where the energy of photons is much less than energy of excitation of atom. Specific feature of considered radiation is that the phases of oscillations of atoms correlate between themselves as oscillations are excited by the same channeling particle. Therefore radiation should be considered as coherent.

\section{Dynamics of atoms in a crystal lattice}
Let us consider very simplified model of a crystal lattice which allows, nevertheless, to find out the basic properties of radiation of the excited atomic chains. We will consider a crystal as a two-dimensional lattice of right-angled structure. Let the relativistic, positively charged particle having a charge $q _ {1} $ enter in a crystal between two atomic chains under small angle to the crystal axes. We consider the energy of the particle much above the energy loss on radiation and interaction with the crystal. Then the velocity of the charged particle in the crystal remains to be constant and equal to the initial velocity $ \vec V $.

Now we are interested in oscillations of the atoms at the lattice points which is caused by collision with the channeling particle, and in the radiation related to these oscillations. In spite of the fact that the atom, as a whole, is neutral, only part of its electrons are bound to the atom. The peripheral electrons belong to the whole crystal. Colliding with an atom, a channeling particle interacts with the  screened potential of the atom nucleus, disturbing it from the equilibrium. Oscillations of a nucleus with interior electrons causes time-dependent polarization of atom and hence electromagnetic radiation. At calculation of interaction of a particle with individual atom we assume that the channeling particle moves parallel to atomic chains with a constant velocity. Rather slow oscillations of a particle between atomic planes we will take into account by the distance between a particle and nucleus of atoms, which is slowly changing along a trajectory of channeling particle. The model of a crystal lattice considered here has been used in \cite {Gogolev} for calculation of polarization of atomic chains induced by the  channeling particle.

\begin{figure} [htbp]
\centerline {\includegraphics[width=5in]{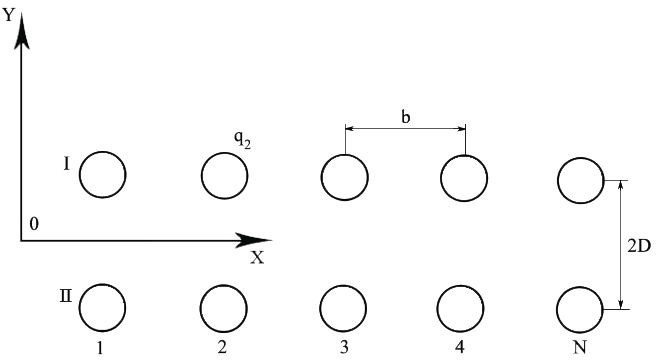}}
\caption {Model of a two-dimensional crystal lattice.}
\label{fig1}
\end{figure}

Let the axis $OX$ of a coordinate system be parallel to the atomic chains and lying  in the middle between the neighboring atomic chains of a crystal (fig. \ref {fig1}). We denote the distance between the next atoms lying in a chain by $b $ and the distance between the atomic chains by $2D $.

Then the atom with number $i$ of the first chain has coordinates $ (x_i, D) $, and the opposite atom of the second chain has coordinates $ (x_i, - D) $. We choose the origin of coordinates system so that $x_0=0$.

The field of a relativistic channeling particle in the  laboratory reference system is concentrated basically in a plane which is perpendicular to a direction of motion \cite {Landau}.

Hence, the particle interacts effectively with the atom during a short interval of time when the atom is in the vicinity of this plane. One can expect that this interaction will be reduced to the momentum transfer from the particle to atom in a direction orthogonal to the velocity of channeling particle. In order to prove it, we find $x$ and $y$ components of the momentum transmitted to the atom of a lattice. The electric field of a charged particle moving with constant velocity  is set by the  formula \cite {Landau}:
\begin{equation} \label{C9}
\vec {E} = \frac {q _ {1} \vec n} {R ^ {2}} \frac {(1-\beta ^ {2})} {\left (1-\beta ^ {2} \sin ^ {2} \psi\right) ^ {3/2}},
\end{equation}
where  $ \vec R $ is a vector pointing from the charged particle to the atom, $ \beta =V/c $, $c$ is the velocity of light, $\psi$ is an angle between the velocity of the moving particle and vector $ \vec R $. Vector $ \vec n $ is a unit vector of direction defined by $ \vec n =\vec R/R $. The angle $ \psi $ depends on parameter $a$ which is the shortest distance between the trajectory of channeling particle and a nucleus of atom.
\begin{equation} \label{psi}
\sin\psi =\frac {a} {R} = \frac {a} {\sqrt {a ^ {2} +V ^ {2} t ^ {2}}}.
\end{equation}
Time $t $ is equal to zero when the particle is at the minimal distance from a nucleus. The atom is acted on by the field of  moving positively charged particle with the Coulomb's  force
\begin{equation} \label{A1}
\vec {F} = \vec {E} q _ {2}.
\end{equation}
where $q_2$ is an effective charge of a nucleus of atom which is screened by the interior electrons. It is obvious that atoms of a crystal lattice are neutral, but a channeling particle approaching the lattice point of a crystal appears to be inside the atom and interacts therefore with the nucleus partially screened by the interior electrons. Consequently, an effective charge  $q_2$ is less than the charge of a nucleus.

One can see from the formula (\ref {C9}) that the field of an ultrarelativistic particle ($ \beta \sim 1$) has important quantity in a vicinity $ \sin\psi\sim 1$ or more exactly $ \cos\psi\sim \gamma ^ {-1} $, where $ \gamma ^ {-1} = \sqrt {1-\beta^2} $ is the relativistic factor. Then, it follows from  (\ref {psi})  that interaction of a particle with a nucleus occurs during time:
\begin{equation}\label{time}
t\sim \frac{a}{V\gamma}\approx\frac{a}{c\gamma}.
\end{equation}
Integrating coordinate components of formula (\ref {A1}) over time $t\in [-\infty, \infty] $, we find components of momentum $P _ {x} $ and $P _ {y} $ which is transmitted to atom of chain along  the axes $OX $ and $OY $ respectively:
\begin{equation}
P_{x} =\frac{q_{1} q_{2}}{aV} \sqrt{{1}-\beta ^{2}},
\end{equation}
\begin{equation}\label{C4}
P_{y} =\frac{q_{1} q_{2} }{aV}.
\end{equation}
It follows from the ratio
$$
\frac {P _ {y}} {P _ {x}} = \frac {1} {\sqrt {1-\beta ^ {2}}} = \gamma,
$$
that the momentum transferred to the atom in $Y$  direction is much greater then that in $X$ direction. Hence, in case of an ultrarelativistic particle  motion  of the atom  along the axis $OX $ can be neglected and only oscillations along the axis $OY$ can be considered. The passing particle excites harmonic oscillations of the atom at frequency $ \omega $ which is defined by crystal properties. For the moment we consider oscillations of atom as free ones, attenuation of these oscillations we shall take into account later.
The law of motion of the $i$-th atom of the first chain can be written as:
\begin{equation} \label{C5}
y^1_i =A^1_i\sin \omega\left (t-\frac {ib} {V} \right) +D, \quad t\ge\frac {ib} {V}.
\end{equation}
Accordingly, the  law of motion for the atom of the second chain will be written as:
\begin{equation} \label{C6}
y^2_i =-A^2_i\sin \omega\left (t-\frac {ib} {V} \right) - D, \quad t\ge\frac {ib} {V},
\end{equation}
where $A_i^j \, \, (j=1, 2) $ is a constant defined by initial conditions. The origin of coordinates is accepted so that $t=0$ when the channeling particle hits the atom number $i=0$. The amplitude of oscillations is defined by an impulse transmitted to atom. It is obvious that the  laws of motion (\ref {C5}) and (\ref {C6}) should be completed by equation
$$ y_i^j=0, \quad t <\frac {ib} {V}. $$
Now, in order to calculate the amplitude of oscillations we suppose that the particle transmits part of its momentum to the atom in no time. Such assumption is valid, if effective time of interaction of a particle with an atom is much less than period of its own oscillations. From the formula (\ref {time}) follows that this condition takes form
$$\frac {\omega a} {c} \ll\gamma,$$
which obviously is fulfilled for a nonrelativistic atom as $ \omega a $ is of the same order of magnitude as the velocity of atom oscillating with amplitude $a $. The amplitude of oscillations we find from initial condition that the amplitude of momentum of the atom $m\omega A_i^j $ is equal to momentum (\ref {C4}) accepted from the particle. As a result we have:
\begin{equation} \label{A8}
A_i^j = \frac {q_1q_2} {m\omega Va_i^j},
\end{equation}
where $a_i^j $ is the shortest distance between the particle and $i$-th atom of chain number $j $.
As the velocity of atom is the nonrelativistic, one can use dipole approximation at calculation of radiation of an oscillating charge. We will combine two atoms which are in different chains opposite each other and are simultaneously excited by a passing particle into one dipole. The dipole moment of obtained system is equal to $ \vec p_i = (2x_iq_1, \, p _ {yi}, \, 0) $, where
\begin{equation} \label{C7}
p _ {yi} (t) =q_2 (A ^ {1} _ {i}-A ^ {2} _ {i}) \sin\omega\left (t-\frac {ib} {V} \right).
\end{equation}
Let us find dependence of  parameter $a_i^j $ on coordinate $x$. Motion of a channeling particle in case of small amplitude of oscillations (it means the amplitude of oscillations in  $Y$ axis direction is much less than interatomic distance $2D$) to a good approximation can be described by the equations:
\bea
y_c=K\cos (\Omega t + \phi_0), \quad x_c=Vt, \nn
\eea
where $y_c $ and $x_c $ are particle coordinates, $K $ and $ \phi_0$ is amplitude and  initial phase. The initial phase can be putted equal to zero by corresponding choice of the origin of $X$ axis\footnote{We have already defined the coordinate origin just after Eq. (\ref{C6}), bounding it to some specific atom. Now we can shift the origin by integer number of distance $b$.  Note that $b\ll V/\Omega$. }. Then the equation of trajectory of a channeling particle takes the form:
$$ y_c=K\cos \frac {\Omega x_c} {V}. $$
Accordingly,  parameters $a_i^j$ for the atoms having  coordinates $x_i=ib $, are equal to
\begin{equation}
a_i^1=D-K\cos\frac {\Omega x_i} {V}, \quad a_i^2=D+K\cos\frac {\Omega x_i} {V}.
\end{equation}
Substituting these parameters in formulas (\ref {A8}) and (\ref {C7}), we  obtain amplitude of oscillations of $i $-th dipole (\ref {C7})
\begin{equation} \label{amp}
A_i=q_2 (A ^ {1} _ {i}-A ^ {2} _ {i}) = \frac {2q_1q_2^2K} {mVD^2\omega} \cos \frac {\Omega x_i} {V},
\end{equation}
And the net result for the dipole moment generated by two atoms of the neighbor chains will be written as:
\begin{equation} \label{C8}
p _ {yi} (x_i, t) =A_i\sin\omega\left (t-\frac {x_i} {V} \right), \quad p _ {xi} (x_i) =2x_iq_1.
\end{equation}
In principle, the form of expression  (\ref {C8}) was obvious from the beginning. All previous reasoning has been directed to calculation of amplitude of oscillations (\ref {amp}). At this point we can correct the primitive model of a crystal lattice to some extend setting the amplitude $A_i $ from some empirical data. For example, we have  not set the quantity of screened charge of nucleus $q_2$, have not considered thermal oscillations of atom etc. Some of such corrections can be made by inserting of corresponding coefficients in  formula (\ref {amp}).

The most essential improvement we are going to make at this step --  is to take into account the attenuation of atom oscillations. The main cause of attenuations is the transmission of the energy of oscillations to surrounding atoms and to a lattice as a whole. In terms of quantum mechanics this process is described by radiation of phonons. By assuming that the energy of phonon is much less than the energy of oscillations we can consider attenuation of oscillations by putting an exponential factor into amplitude of oscillations. With account of attenuation the equation for the dipole moment (\ref {C8}) becomes:
\begin{equation}\label{C12}
p_{yi}(x_i,t)=\begin{cases}A_ie^{-\alpha(t-\frac{x_i}{V})}\sin\omega\left(t-\displaystyle\frac{x_i}{V}\right),\quad t>\displaystyle\frac{x_i}{V},\\0,\,t\leq\displaystyle\frac{x_i}{V},\end{cases}
\end{equation}
where $ \alpha $ is the attenuation coefficient. We will ignore the constant component of the dipole moment $p _ {xi} $  as the radiation field is proportional to the second derivative on time from the dipole moment.

\section{Radiation of the excited chains of atoms}

\begin{figure} [h]
\centerline{\includegraphics[width=3in]{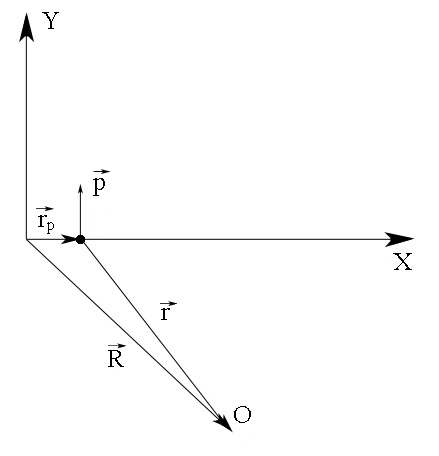}}
\caption {System of coordinates.}
\label{figure2}
\end{figure}

Electric field of radiation of the dipole moment is defined by the formula \cite {Landau}:
\begin {equation} \label {field}
\vec {E} _ {i} (\vec {r}, t) = \frac {1} {rc ^ {2}} \left [\vec {n} \left [\vec {n}, \ddot {\vec {p}} _ {i} (t ') \right] \right],
\end {equation}
where the unit vector $ \vec n $ is defined by equality $ \vec n =\vec r/r $, $ \vec r $ is the vector pointing from the  dipole to the point for which the vector $ \vec E_i $ is written down, $r = |\vec r | $, $ \vec p_i $ is to be evaluated at the retarded time $t ' $
$$ t ' =t-\frac rc. $$
Further we assume that the distance between the crystal and the observer is much greater than sizes of the crystal. Then the vector $ \vec n $ and distance $r $ in Eq. (\ref {field}) are constant and do not depend on $x $. Dependence on $x $ should be considered only in the phase of an electromagnetic wave, namely in $t ' $. We denote the vector from the coordinates origin to the point of observation by $ \vec R $, and radius-vector of the dipole moment by $ \vec r_p $  as shown in Fig. \ref {figure2}.
As $R\gg r_p $, it is possible to write approximately:
\begin {equation}
\begin {array} {l}
r =\sqrt {(\vec R-\vec r_p) ^2} \approx R - (\vec {n} \vec {r_p}),
\end {array}
\end {equation}
where $ \vec {n} = \vec R/R $. Accordingly, for the retarded time  $t ' $ we have:
\begin {equation} \label {retard}
t ' =t-\frac {R-xn_x} {c}.
\end {equation}
Let us find the components of vector $ \vec E (t) $ in a spherical system of coordinates shown in Fig. \ref {figure3}. Direction of the unit vector $\vec n$ is defined by the polar angle $ \theta $ and azimuthal  angle $\varphi$. The Cartesian coordinates  of the unit vectors of spherical  coordinate system can be written as:
\begin{figure}[h]
\centerline{\includegraphics[width=3in]{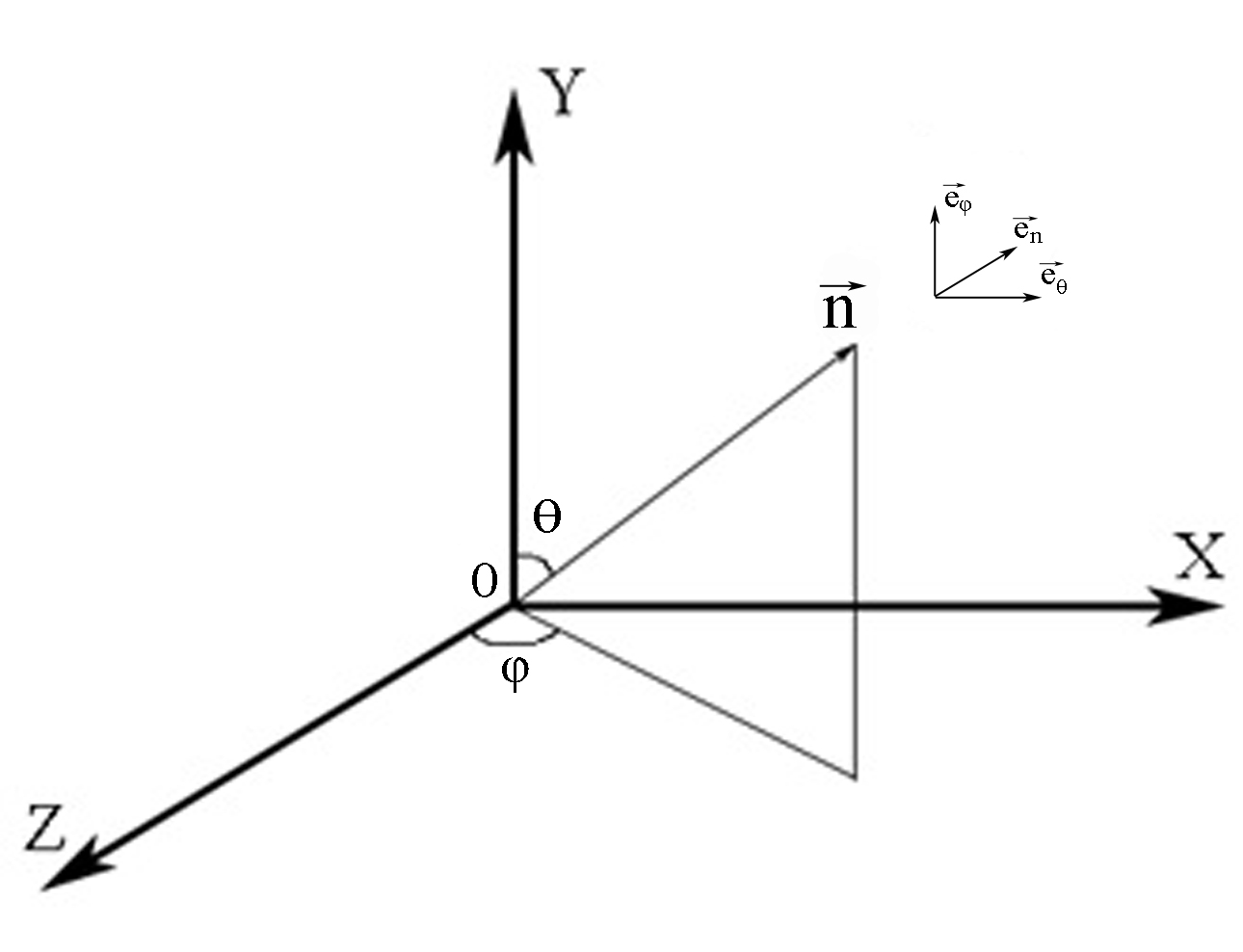}}
\caption{ Spherical system of coordinates.}\label{figure3}
\end{figure}
\bea\label{orts}
\vec{n}&=&\left(\sin\theta \sin\varphi, \cos\theta, \sin\theta \cos\varphi\right),
\\
\vec e_\theta&=&(\cos\theta\sin\varphi, -\sin\theta, \cos\theta\cos\varphi),
\\
\vec e_\varphi&=&(\cos\varphi, 0, -\sin\varphi).
\eea
Then it follows from Eq. (\ref {field}) that:
\begin{equation} \label {eteta}
\vec {E_i} (\vec R, t) = \frac {\vec {e} _ {\theta}} {Rc ^ {2}} \ddot p _ {yi} (t ') \sin\theta.
\end{equation}
Taking twice derivative from Eq. (\ref {C12}) with respect to time, we obtain
\begin{equation} \label{pddot}
\ddot {p} _ {yi} (x_i, t ') = \begin{cases} A_ie ^ {\displaystyle-\alpha\tau} \left [(\alpha^2-\omega^2) \sin\omega\tau-2\alpha\omega\cos\omega
\tau\right], \quad t '> \dfrac {x_i} {V}, \\
0, \quad t '\leq\displaystyle\frac {x_i} {V}, \end{cases}
\end{equation}
where $$\tau=t '-\frac {x_i} {V} =t-\frac Rc-\frac {x_i} {V} (1-\beta n_x), \quad \beta =\frac Vc. $$
Let the atomic chain consist of large, but finite number of atoms $N $. The net field of radiation is the discrete sum of fields of individual atoms of the crystal
\begin{equation}
\vec E (\vec R, t) = \sum _ {i = {0}} ^ {N} \vec E _ {i} (\vec R, t).
\end{equation}
As the distance between the atoms is much less than the wavelength of radiation, it is possible to turn to continuous distribution of the dipole moment along the $X$-axis, and replace the sum  by integral. There is one dipole per interval of length $b$ in atomic chain, hence, the quantity $ \varepsilon (x, t) =E _ {i\theta} (\vec R, t)/b $ has the meaning of the  field generated by unit length of the atomic chain. Substituting into this expression the field of Eq. (\ref {eteta}) and taking into account Eq. (\ref {pddot}), we obtain $ \varepsilon (x, t) $ for a time interval where it is not equal to zero:
\begin{equation} \label{epsylon1}
\varepsilon (x, t) =A_0e ^ {-\alpha\tau} \cos\frac {\Omega x} {V} \left [(\alpha^2-\omega^2) \sin\omega\tau-2\alpha\omega\cos\omega
\tau\right],
\end{equation}
here
\begin{equation} \label{amplit}
A_0 =\frac {2q_1q_2^2K\sin\theta} {mVD^2\omega bRc^2},
\quad \tau=t-\frac Rc-\frac {x} {V} (1-\beta n_x).
\end{equation}
Trigonometric functions in the square brackets can be combined, using the phase $\psi$:
\begin{equation} \label{psi1}
\sin\psi\omega =\frac {2\alpha\omega} {\alpha^2 +\omega^2}, \quad\cos\psi\omega =\frac {\alpha^2-\omega^2} {\alpha^2 +\omega^2}.
\end{equation}
As a result we get:
\begin{equation} \label{epsylon2}
\varepsilon (x, t) =A_0 (\alpha^2 +\omega^2) e ^ {-\alpha\tau} \cos\frac {\Omega x} {V} \sin\omega (\tau-\psi).
\end{equation}
Integrating this expression over $x $, we  find the field generated by pair of atomic chains
\begin{equation} \label{eott}
E (t) _ \theta =\int\limits _ {x_1} ^ {x_2} \varepsilon (x, t) \, dx.
\end{equation}
The lower limit $x_1$ of integral  is coordinate of the crystal origin. The upper limit $x_2$ is a time function.

While the channeling the particle moves in a crystal, the upper limit coincides with coordinate of the particle according to Eq.  (\ref {pddot}). After the particle leaves the crystal, radiation is generated by all the chain of atoms, and $x_2$ coincides with coordinate of the end of chain $x _ {20} $:
\begin{equation} \label{xupper}
x_2 (t) = \begin{cases} (ct-R) \dfrac {\beta} {(1-\beta n_x)}, \quad t\leq\dfrac Rc +\dfrac {x} {V} (1-\beta n_x), \\
 x _ {20}, \quad t> \dfrac Rc +\dfrac {x} {V} (1-\beta n_x).
\end{cases}
\end{equation}
Integration in Eq.  (\ref {eott}) gives
\bea\label{eott2}
E_\theta(t)=\frac{1}{2k}A_0(\alpha^2+\omega^2)&e^{-\alpha(t_0-kx)}&\left(\frac{\alpha\sin(\omega t_1+kx\Delta_-)-\Delta_-\cos(\omega t_1+kx\Delta_-) }{\alpha^2+\Delta_-^2}\right.\nonumber\\
&+&\left.\left.\frac{\alpha\sin(\omega t_1-kx\Delta_+)+\Delta_+\cos(\omega t_1+kx\Delta_+)}{\alpha^2+\Delta_+^2} \right)\right|^{x_2}_{x_1},
\eea
where	
\begin{equation} \label{delta}
t_0=t-\frac Rc, \quad t_1=t-\frac Rc-\psi, \quad k =\frac 1V (1-\beta n_x), \quad \Delta _ {\pm} = \Omega '\pm\omega.
\end{equation}
We see from this expression that radiation is generated on two frequencies, namely on frequency of atoms oscillations $\omega$ and the frequency of oscillations of channeling particle which is shifted by the Doppler effect:
$$\Omega'=\frac{\Omega}{1-\beta n_x}.$$
Eq.  (\ref {eott2}) can be written as:
\bea\label{eott3}
E_\theta (t) =e ^ {-\alpha t_0} \left [F (x_2, t) e ^ {\alpha kx_2}-F (x_1, t) e ^ {\alpha kx_1} \right],
\eea
where $F (x, t) $ is periodic function of time. Considering exponential factor in last formula, we see that while
$$
 t\leq\dfrac Rc +\dfrac {x} {V} (1-\beta n_x),
 $$
the amplitude of the electric field of radiation is growing accordingly to
\bea\
E_\theta (t) =F (x_2, t) e ^ {\alpha kx_2}-F (x_1, t) e ^ {\alpha (kx_1-t_0)},\nn
\eea
and approaching a constant level  if the crystal is long enough. It means that the time it takes for the particle to cross the crystal is much greater then the lifetime of oscillating atoms $1/\alpha$. We shall refer such a crystal as a ``long'' one. After a particle leaves the crystal, the amplitude of  field exponentially decreases.
Let us next neglect the boundary effects and consider only radiation with constant amplitude of the field. In other words, we consider only first term in Eq. (\ref {eott3}) where $x_2=t_0/k $. This is the main term in case of a ``long crystal''. Then the formula for a field of radiation becomes
\bea\label{eott5}
E_\theta(t)&=&\frac{1}{2k}A_0(\alpha^2+\omega^2)\left\{\sin\Omega't\left[\frac{\alpha\cos\omega\psi-\Delta_-\sin\omega\psi}{\alpha^2+\Delta_-^2}-\frac{\alpha\cos\omega\psi+\Delta_+\sin\omega\psi}{\alpha^2+\Delta_+^2}\right]\right.\nonumber\\
&+&\left.\cos\Omega't\left[\frac{-\alpha\sin\omega\psi-\Delta_-\cos\omega\psi}{\alpha^2+\Delta_-^2}-\frac{\alpha\sin\omega\psi-\Delta_+\cos\omega\psi}{\alpha^2+\Delta_+^2}\right]\right\}.
\eea
As we see in this case the radiation is generated only on frequency $ \Omega ' $.
Let us find angular distribution of intensity of radiation. Intensity of radiation $dI $ in an element of space angle $do $ is defined by the formula \cite {Landau}
\begin{equation} \label {angular1}
\frac {dI} {do} = \frac {cR^2E^2} {4\pi}.
\end{equation}
Averaging expression (\ref {eott5}) over the period of radiation and substituting in (\ref{angular1}) we obtain
\begin{equation}\label{angular2}
\frac{dI}{do}=\frac{q_1^2q_2^4 K^2\sin^2\theta}{2\pi m^2b^2c^3D^4(1-\beta n_x)^2}\frac{4\Omega'^2\alpha^2+(\alpha^2+\omega^2)^2}{(\alpha^2-\omega^2+\Omega'^2)^2+4\alpha^2\omega^2}.
\end{equation}
From last formula follows that dependence of intensity of radiation on frequencies $ \Omega ' $ and $ \omega $ has resonant nature. At a reasonably small $ \alpha $ the resonance occurs at $ \Omega ' = \omega $ or
\begin{equation} \label {C10}
\frac {\Omega} {1-\beta n_x} = \omega.
\end{equation}
If the channeling particle is ultrarelativistic one, sufficient  wide range of values of the relation $\omega/\Omega '$ satisfies the resonance condition (\ref {C10}):
$$\frac {1} {2} <\frac {\omega} {\Omega '} <2\gamma^2. $$

\section{Conclusion}
In this work the basic properties of radiation of the atomic chains excited by a channeling particle are considered. On a simplified model of a crystal lattice it is shown that this radiation is generated on two frequencies, namely on the  frequency of  atom oscillations, and on the frequency of oscillations of a channeling particle, shifted by Doppler effect. In case of a reasonably long crystal the radiation with frequency of oscillations of a channeling particle prevails. Occurrence of the factor $ (1-\beta n_x) $ in denominator of the formula for angular distribution of intensity of radiation (\ref {angular2}) results in relativistic effect -- the basic part of radiation is generated into a cone around the  direction of velocity of a relativistic  channeling particle. Hence, the frequency of radiation and angular distribution are similar to corresponding characteristics of radiation of a channeling particle. This can rise a problem of experimental separation of one form of radiation from another one and a problem of identification of considered radiation.

The interesting effect showing that a set of not relativistic oscillators at rest or oscillating atoms produces radiation with the properties which are specific for radiation of the relativistic charged particle, is obviously a result of the fact that oscillations of atoms are coherent. Because phases of their oscillation are tuned by the channeling particle. The interference of fields of radiation of individual atoms gives  the properties of the net radiation, described above.

\subsection*{Acknowledgment}

The authors would like to thank Professor V.G.~Tyuterev for valuable discussions.
This work has been supported by the grant for LRSS, project No 3558.2010.2

\end{document}